\documentclass[11pt, a4paper]{article}

\usepackage[utf8]{inputenc}
\usepackage[T1]{fontenc}
\usepackage{amsmath, amssymb}
\usepackage{geometry}
\usepackage{graphicx}
\usepackage{float}
\usepackage{tikz}
\usepackage{xcolor}
\usepackage{titlesec}
\usepackage[hidelinks]{hyperref}
\usepackage{booktabs}
\usepackage{multirow}

\usetikzlibrary{arrows.meta, positioning, fit, backgrounds, calc}

\pgfdeclarelayer{background}
\pgfsetlayers{background,main}

\definecolor{azul}{RGB}{31,119,180}
\definecolor{verde}{RGB}{44,160,44}
\definecolor{cinza}{RGB}{100,100,100}
\definecolor{fundocaixa}{RGB}{240,248,255}

\titleformat{\section}{\large\bfseries}{\thesection.}{0.5em}{}
\titleformat{\subsection}{\normalsize\bfseries}{\thesubsection}{0.5em}{}
\titlespacing*{\section}{0pt}{1.8ex plus 0.5ex minus 0.2ex}{1ex plus 0.2ex}
\titlespacing*{\subsection}{0pt}{1.4ex plus 0.3ex minus 0.1ex}{0.6ex plus 0.1ex}

\begin{document}

\begin{center}
{\LARGE\bfseries Sparse Principal Component Analysis via Wavelets for Distributed Data}
\end{center}

\begin{center}
    Giovanni Barbosa Herrero
    \footnote[1]{IC, UNICAMP -- g281775@dac.unicamp.br}

    Rodney Vasconcelos Fonseca
    \footnote[2]{IME, UFBA -- rodneyfonseca@ufba.br}

    Alu\'isio Pinheiro
    \footnote[3]{IMECC, UNICAMP -- pinheiro@unicamp.br}
\end{center}

\begin{abstract}
The large volume of data and concerns about data privacy have motivated the development of techniques for distributed data, a problem also known as federated learning. In this scenario, sub-samples of the data are divided across different machines, and statistics must be computed over that data without direct access to the full sample. Johnstone \& Lu (2009, JASA) show that principal component analysis (PCA) is statistically inconsistent in the high-dimensional regime, and propose a way to recover consistency through wavelet-based sparsification and variable selection. Fan et al.\ (2019, AoS) show a way to perform this same estimation, specifically, to estimate the eigenspace that would be obtained if all the data were pooled together, even though it remains effectively distributed, without addressing the high-dimensional regime. This work incorporates the wavelet-based sparsification of Johnstone \& Lu (2009) into the distributed PCA framework of Fan et al.\ (2019), aiming to reduce communication cost without compromising the quality of the eigenspace estimation. Simulations across $d \in [52, 5000]$ show that the proposed method overtakes Fan et al.\ (2019) in estimation error beyond a clear dimensional threshold ($d \geq 152$ for $\lambda=25$, $d \geq 252$ for $\lambda=50$), while transmitting systematically fewer coefficients throughout the entire range studied. This study was financed by the S\~ao Paulo Research Foundation (FAPESP), Brazil. Process Number \#2023/02538-0 and Number \#2025/21250-2.
\end{abstract}


\section{Introduction}

The last decade has witnessed a technological revolution marked by the vertiginous growth in the volume and complexity of collected data, a phenomenon popularized under the term \textit{big data} and that drove the consolidation of Data Science as a central discipline in contemporary statistics. A direct consequence of this growth was the emergence of the so-called high-dimensional regime, in which the number of observed variables $p$ becomes comparable to, or even larger than, the number of available samples $N$, contradicting the classical asymptotic premise of fixed $p$ and $N \to \infty$.

In this scenario, classical statistical techniques have proven inadequate. In particular, the conventional Principal Component Analysis (PCA) method loses its usual asymptotic properties: Johnstone \& Lu (2009) showed that the eigenvectors estimated by sample PCA can be statistically inconsistent when $p/N$ does not approach zero, that is, the classical estimator does not converge to the true population eigenvector as the dimension grows. To circumvent this problem, the authors assume the existence of a basis in which the data admit a sparse representation, that is, a basis in which only a small subset of coordinates concentrates most of the relevant information. This sparsity assumption enables a prior variable-selection step, which restricts PCA to that reduced subset of coordinates before estimating the eigenvectors, allowing consistency of the estimator to be recovered even in high dimension.

As a basis for this sparse representation, Johnstone \& Lu (2009) use wavelets. This choice is justified by the ability of wavelets to provide representations that are localized both in space and in frequency: unlike classical bases, such as the Fourier basis, whose elements have global support, the functions of a wavelet basis have compact support and concentrate the signal's energy in a few coefficients of relevant magnitude, with the remaining ones close to zero. This property makes wavelets particularly efficient at promoting sparse representations in signals with local structure --- peaks, discontinuities, or concentrated variations --- which explains their widespread use in compression and denoising problems (Morettin, Pinheiro \& Vidakovic, 2017).

We also use wavelets as the sparsification basis in this work, these functions being central mathematical objects of our research.

In parallel, present-day data is increasingly stored in a distributed fashion, and statistics must be computed over that data, a problem known as Federated Learning. Consider a dataset $\{x_i\}_{i=1}^N$ of vectors $x_i \in \mathbb{R}^p$, distributed across $M$ separate machines, each with $n = N/M$ samples. These machines are connected only to a central server, but the raw data cannot be sent to that server, only statistics computed locally from the data. This scenario stands in contrast to the classical paradigm, in which the whole dataset is centralized and available in a single location for the computation of statistics, and typically arises when one wishes to preserve data privacy, or when the dataset is too large to fit on a single machine. Sending all observations to a central server may, moreover, be prohibitive due to the required bandwidth. Specifically for PCA, one wishes to estimate the principal eigenspace that would be obtained if all the data were pooled together in a single location, even though it remains effectively distributed. Fan et al.\ (2019) show a way to perform this estimation, without, however, addressing the high-dimensional regime. Their analysis implicitly assumes that $p$ remains controlled relative to the local sample size.

Our research investigates precisely the intersection of these two regimes: high dimension \textit{and} distributed data simultaneously. This combined scenario is increasingly common in practice, for example, in hospital networks that need to aggregate patient statistics on a central server without transmitting individual data, due to regulatory compliance requirements, or in wearable devices (\textit{smartwatches}), which continuously collect high-dimensional signals but have limited communication capacity. The main goal of this work is, therefore, to study the application of wavelet-based methods to the estimation of sparse principal components for distributed data, proposing a fusion between the wavelet-based sparsity approach of Johnstone \& Lu (2009) and the distributed framework of Fan et al.\ (2019); the details of this combination are discussed in the Methodology section.

\section{Methodology}

We begin by briefly describing the algorithms of Fan et al.\ (2019) and of Johnstone \& Lu (2009), respectively. The notation used here follows that of the original papers and will be used throughout the description of our algorithm, with small modifications to avoid ambiguity.

\subsection{Fan et al.\ (2019)}

Suppose a horizontally distributed system (that is, the data is partitioned by samples: each machine stores a complete subset of observations, rather than a subset of variables) with $N$ i.i.d.\ sub-Gaussian random samples $\{\mathbf{X}_i\}_{i=1}^N \subseteq \mathbb{R}^d$, with $\mathbb{E}\mathbf{X}_1 = \mathbf{0}$ and covariance matrix
\[
    \mathbb{E}(\mathbf{X}_1\mathbf{X}_1^\top) = \boldsymbol{\Sigma} \in \mathbb{R}^{d \times d},
\]
and $m$ machines, each containing $n$ samples, such that $N = nm$. Assume that samples are i.i.d.\ within each machine and, for simplicity, i.i.d.\ across machines. The algorithm proposed by Fan et al.\ (2019) estimates $\mathrm{Col}(\mathbf{V}_K)$, the column space spanned by the top $K$ eigenvectors of $\boldsymbol{\Sigma}$, and works as follows: for each machine $\ell \in \{1, \ldots, m\}$, compute the top $K$ eigenvectors $\hat{\mathbf{V}}_K^{(\ell)} \in \mathbb{R}^{d \times K}$ of the local sample covariance matrix
\[
    \hat{\boldsymbol{\Sigma}}^{(\ell)} = (1/n)\sum_{i=1}^n \mathbf{X}_i^{(\ell)} \mathbf{X}_i^{(\ell)\top},
\]
and transmit $\hat{\mathbf{V}}_K^{(\ell)}$ to the central server. At the central server, compute the average of the projection matrices
\[
    \tilde{\boldsymbol{\Sigma}} = (1/m)\sum_{\ell=1}^m \hat{\mathbf{V}}_K^{(\ell)} \hat{\mathbf{V}}_K^{(\ell)\top}
\]
and take its top $K$ eigenvectors as the final estimator $\tilde{\mathbf{V}}_K$. The communication cost is of order $O(mKd)$, where, in most cases, $K = o(\min(n,d))$. Statistical error is measured by
\[
    \rho(\tilde{\mathbf{V}}_K, \mathbf{V}_K) = \|\tilde{\mathbf{V}}_K\tilde{\mathbf{V}}_K^\top - \mathbf{V}_K\mathbf{V}_K^\top\|_F,
\]
the Frobenius norm of the difference between the projection matrices. Fan et al.\ (2019) show that $\tilde{\mathbf{V}}_K$ enjoys the same statistical error rate as PCA computed on the full sample.

\subsection{Johnstone \& Lu (2009)}

Johnstone \& Lu (2009) show that PCA in high-dimensional regimes is inconsistent, and propose an algorithm based on prior dimensionality reduction through variable selection. Precisely, the standard PCA estimator is consistent if and only if $d/n \to 0$ (Theorem 1 in Johnstone \& Lu, 2009). To recover consistency, the authors suggest working in a basis in which the signals have a sparse representation, performing variable selection before PCA. In this work, we will use wavelets, as in Johnstone \& Lu.

Suppose $\{\mathbf{x}_i,\, i = 1,\ldots,n\}$ is a dataset of $n$ observations on $d$ variables, with $d$ comparable to $n$. Standard PCA on this dataset yields inconsistent results. To recover consistency, Johnstone \& Lu propose the following algorithm:

\begin{enumerate}
    \item \textit{Compute basis coefficients.} Given a basis $\{\mathbf{e}_\nu\}$ for $\mathbb{R}^d$ in which the data have a sparse representation, compute coordinates $x_{i\nu} = \langle \mathbf{x}_i, \mathbf{e}_\nu \rangle$ for each $\mathbf{x}_i$.

    \item \textit{Subset.} Compute the sample variances $\hat{s}^2_\nu = \widehat{\mathrm{var}}(x_{i\nu})$. Let $\hat{I} \subset \{1,\ldots,d\}$ denote the set of indices $\nu$ corresponding to the largest $\hat{k}$ variances.

    \item \textit{Reduced PCA.} Apply standard PCA to the reduced dataset $\{x_{i\nu},\, \nu \in \hat{I},\, i = 1,\ldots,n\}$ on the selected $\hat{k}$-dimensional subset, obtaining eigenvectors $\hat{\mathbf{v}}^j$, $j = 1,\ldots,\hat{k}$.

    \item \textit{Thresholding.} Filter out noise in the estimated eigenvectors by \textit{hard thresholding}:
    \[
        \hat{v}^j_\nu \leftarrow \eta_H(\hat{v}^j_\nu,\, \delta_j).
    \]

    \item \textit{Reconstruction.} Return to the original signal domain using the basis $\{e_\nu\}$ and set
    \[
        \hat{v}_j(t_l) = \sum_{\nu \in \hat{I}} \hat{v}^j_\nu\, e_\nu(t_l).
    \]
\end{enumerate}

\noindent\textit{Remark on Step 5.} The notation $t_l = l/d$ reflects the fact that Johnstone \& Lu (2009) originally consider data collected in the time domain, in which each observation $\mathbf{x}_i$ is a discretized function evaluated at $d$ equally spaced points, $t_l$, $l=1,\ldots,d$. The reconstruction step returns the estimated eigenvector to that time-domain representation.

\subsection{Our proposal}

We consider the distributed setting of Fan et al.\ (2019): a dataset of $N = mn$ i.i.d.\ Gaussian samples $\{\mathbf{X}_i\}_{i=1}^N \subseteq \mathbb{R}^d$ with $\mathbb{E}\mathbf{X}_1 = \mathbf{0}$ and population covariance matrix $\boldsymbol{\Sigma}$, distributed across $m$ machines, each storing $n$ observations $\mathbf{X}_1^{(\ell)}, \ldots, \mathbf{X}_n^{(\ell)} \in \mathbb{R}^d$, where $\ell \in \{1,\ldots,m\}$ denotes the index of a specific machine. We further assume a high-dimensional regime in which $d$ is comparable to $n$, which motivates the use of wavelet-based sparsification prior to transmission. The goal is to estimate $\mathrm{Col}(\mathbf{V}_K)$, the top-$K$ eigenspace of $\boldsymbol{\Sigma}$, while reducing the number of coefficients transmitted to the central server.

The following steps are carried out locally on each machine $\ell \in \{1, \ldots, m\}$.

\begin{enumerate}

    \item \textit{Wavelet Transform.} Apply the discrete wavelet transform (DWT) to each row of $\mathbf{X}^{(\ell)}$, in $\mathbb{R}^{n \times d}$, obtaining $\mathbf{X}_w^{(\ell)}$, also in $\mathbb{R}^{n \times d}$. The subscript $w$ indicates that quantities are in the wavelet domain and will be kept throughout the text.

    \item \textit{Variable selection.} We use the method described in Section 4.2(b) of Johnstone \& Lu (2009) to perform variable selection on $\mathbf{X}_w^{(\ell)}$. Compute the sample variances
    \[
        \hat{s}^2_w = n^{-1}\sum_{i=1}^n x_{iw}^2.
    \]
    Under the assumption that the noise level is the same across coordinates, estimate it by $\hat{\sigma}^2 = \mathrm{median}(\hat{s}^2_w)$. Let $\chi^2_{(n-1),\,\alpha}$ denote the upper-$\alpha$ percentile of the $\chi^2_{(n-1)}$ distribution. Define the excess over the chi-squared \textit{baseline} as
    \[
        \eta^2_{(w)} = \max\!\left\{
        \hat{s}^2_{(w)} - \frac{\hat{\sigma}^2}{n-1}
        \chi^2_{(n-1),\,w/(d+1)},\ 0
        \right\}.
    \]
    For a specified fraction $q(n) \in (0,1)$, define
    \[
        \hat{I}^{(\ell)} = \left\{w : \sum_{w=1}^{\hat{k}^{(\ell)}}
        \eta^2_{(w)} \geq q(n)\sum_{w} \eta^2_{(w)}\right\},
    \]
    where $\hat{k}^{(\ell)}$ is the smallest index $k$ for which the inequality holds. Thus, $|\hat{I}^{(\ell)}| = \hat{k}^{(\ell)}$, and $\hat{I}^{(\ell)}$ contains the indices $w$ that contribute to the left-hand sum of the excess $\eta^2_{(w)}$.

    \item \textit{Reduced PCA.} Let $\mathbf{X}_{\hat{I},w}^{(\ell)} \in \mathbb{R}^{n \times \hat{k}^{(\ell)}}$ be the reduced data matrix, obtained from the original data matrix in the wavelet domain by retaining only the columns indexed by $\hat{I}^{(\ell)}$. Apply standard PCA to the data in $\mathbf{X}_{\hat{I},w}^{(\ell)}$, computing the top $K$ eigenvectors of the local sample covariance matrix $\hat{\boldsymbol{\Sigma}}_{\hat{I},w}^{(\ell)} = (1/n)\mathbf{X}_{\hat{I},w}^{(\ell)\top} \mathbf{X}_{\hat{I},w}^{(\ell)}$, obtaining $\hat{\mathbf{V}}_{K,w}^{(\ell)*} \in \mathbb{R}^{\hat{k}^{(\ell)} \times K}$. The superscript ${}^*$ is used throughout Steps 3--4 to distinguish matrices in $\mathbb{R}^{\hat{k}^{(\ell)} \times K}$ from their counterparts in $\mathbb{R}^{d \times K}$.

    \item \textit{Thresholding.} For each component $j = 1,\ldots,K$ and $w \in \hat{I}^{(\ell)}$, apply a thresholding rule $\eta(\cdot\,;\delta)$ to the $j$-th column of $\hat{\mathbf{V}}_{K,w}^{(\ell)*}$:
    \[
        \hat{v}^{(\ell)*}_{w,j} \leftarrow \eta(\hat{v}^{(\ell)*}_{w,j}\,;\,
        \delta_j^{(\ell)}),
    \]
    where
    \[
        \delta_j^{(\ell)} = c\,\hat{\tau}_j^{(\ell)}\sqrt{2\log \hat{k}^{(\ell)}},
    \]
    \[
        \hat{\tau}_j^{(\ell)} = \mathrm{MAD}\bigl\{\hat{v}^{(\ell)*}_{w,j} : w \in \hat{I}^{(\ell)}\bigr\}\big/0.6745,
    \]
    and $c > 0$ is a tuning parameter specified further ahead, in the ``Simulation design'' subsection. Two thresholding rules $\eta(\cdot\,;\delta)$ are considered in this work: hard thresholding, $\eta_H(v;\delta)$, and the smooth SCAD rule of Kulkarni et al. (2026), $\eta_{\mathrm{SSCAD}}(v;\delta)$, which additionally depends on a shape parameter $a > 1$; both rules, and their respective tuning parameters, are specified in the ``Simulation design'' subsection.

    \item \textit{Embedding and transmission.} Embed $\hat{\mathbf{V}}_{K,w}^{(\ell)*} \in \mathbb{R}^{\hat{k}^{(\ell)} \times K}$, already thresholded, into $\mathbb{R}^{d \times K}$ by setting, for each $j = 1,\ldots,K$ and $w = 1,\ldots,d$,
    \[
        \hat{v}^{(\ell)}_{w,j} =
        \begin{cases}
            \hat{v}^{(\ell)*}_{w,j} & \text{if } w \in \hat{I}^{(\ell)}, \\[4pt]
            0 & \text{otherwise,}
        \end{cases}
    \]
    where $\hat{v}^{(\ell)*}_{w,j}$ denotes the $(w,j)$ entry of $\hat{\mathbf{V}}_{K,w}^{(\ell)*}$ (after thresholding). The resulting matrix $\hat{\mathbf{V}}_{K,w}^{(\ell)} \in \mathbb{R}^{d \times K}$ is sparse, with nonzero rows only at indices in $\hat{I}^{(\ell)}$. Transmit $\hat{\mathbf{V}}_{K,w}^{(\ell)}$ to the central server.

\end{enumerate}

\noindent At the central server:

\begin{enumerate}
    \setcounter{enumi}{5}

    \item \textit{Re-orthonormalization.} Re-orthonormalize the columns of $\hat{\mathbf{V}}_{K,w}^{(\ell)}$, for each $\ell \in \{1,\ldots,m\}$, via the Gram--Schmidt process, restricted to the support $\hat{I}^{(\ell)}$, obtaining $\hat{\mathbf{V}}_{K,w}^{(\ell)} \in \mathbb{R}^{d \times K}$ with orthonormal columns.

    \item \textit{Inverse Wavelet Transform.} Apply the IDWT to each column of $\hat{\mathbf{V}}_{K,w}^{(\ell)}$,
    for each machine $\ell$,
    obtaining $\hat{\mathbf{V}}_K^{(\ell)}$, in $\mathbb{R}^{d \times K}$, in the original domain.

    \item \textit{Aggregation.} Compute the average of the projection matrices
    \[
        \tilde{\boldsymbol{\Sigma}} = \frac{1}{m}\sum_{\ell=1}^m
        \hat{\mathbf{V}}_K^{(\ell)}\hat{\mathbf{V}}_K^{(\ell)\top}.
    \]

    \item \textit{Final PCA.} Take the top $K$ eigenvectors of $\tilde{\boldsymbol{\Sigma}}$ as the final estimator $\tilde{\mathbf{V}}_K \in \mathbb{R}^{d \times K}$.

\end{enumerate}

\subsection{Simulation design}

Our experiment consists of estimating the eigenspace $\mathrm{Col}(\mathbf{V}_K)$, described in the previous section, and comparing the estimation error of our method with the error of the algorithm of Fan et al.\ (2019). We follow the simulation design of Fan et al.\ (2019): samples $\{\mathbf{X}_i\}_{i=1}^N$ are generated i.i.d.\ from $N(\mathbf{0}, \boldsymbol{\Sigma})$, with
\[
    \boldsymbol{\Sigma} = \mathrm{diag}(\lambda, \lambda/2, \lambda/4, 1, \ldots, 1) \in \mathbb{R}^{d \times d},
\]
the \textit{spiked} covariance model of Johnstone (2001), with $K=3$ spiked eigenvalues and population eigenspace $\mathbf{V}_K = (\mathbf{e}_1, \mathbf{e}_2, \mathbf{e}_3)$. We consider signal strengths $\lambda \in \{25, 50\}$, $m=5$ machines, and $n=500$ samples per machine. All reported values are medians over $N_{MC}=100$ paired Monte Carlo replications. The proposed algorithm uses the Daubechies db4 wavelet, decomposition level $J=2$, and selection fraction $q(n) = 0.995$.

Two thresholding rules $\eta(\cdot\,;\delta)$ are applied to the reduced eigenvectors (Step 4 of the ``Our proposal'' subsection), each with its own tuning parameter $c$ entering $\delta_j^{(\ell)} = c\,\hat{\tau}_j^{(\ell)}\sqrt{2\log \hat{k}^{(\ell)}}$:
\begin{itemize}
    \item \textit{Hard thresholding}, $\eta_H(v;\delta)$, with $c=1$, that is, the so-called universal threshold from wavelet theory;
    \item the \textit{smooth SCAD} rule of Kulkarni et al.\ (2026), which additionally depends on a shape parameter $a > 1$ controlling the width of a smooth (raised-cosine) transition zone between full shrinkage and the identity map. We fix $a = 3.7$, the classical value recommended for SCAD-type penalties (Fan \& Li, 2001), and set $c = c(a)$ following the heuristic of Kulkarni et al.\ (2026, Eqs.\ 37--39),
    \[
        c(a) = \sqrt{\frac{a-2}{a-1}},
    \]
    which for $a=3.7$ gives $c(a) \approx 0.79$ -- about $21\%$ smaller than the universal threshold used for hard thresholding.
\end{itemize}
Other values of $c$ and $a$ may be used to regulate the thresholding step. The dimension $d$ is sampled from 52 to 5000, increasing at each step by approximately 50 units, with small adjustments so that every sampled value is a multiple of 4, a condition that guarantees, for the db4 wavelet with $J=2$ decomposition levels under periodic boundary handling, that the dimension in the wavelet domain coincides exactly with $d$. The algorithm was implemented in Python, using PyWavelets (Lee et al., 2019) for the wavelet transforms, NumPy (Harris et al., 2020) and PyTorch (Ansel et al., 2024) for array and GPU-accelerated linear algebra operations, and SciPy (Virtanen et al., 2020) for the $\chi^2$ quantiles used in the variable-selection step. Results were aggregated with pandas (McKinney, 2010) and plotted with Matplotlib (Hunter, 2007).

\subsection{Metrics}

Statistical error is measured by the Frobenius distance between projection matrices,
\[
    \rho = \|\tilde{\mathbf{V}}_K \tilde{\mathbf{V}}_K^\top - \mathbf{V}_K\mathbf{V}_K^\top\|_F,
\]
reported separately as $\rho^{\mathrm{Fan}}$ and $\rho^{\mathrm{Wav}}$ for Fan et al.\ (2019) and for our algorithm, respectively.

We define the communication ratio ($R_{\mathrm{com}}$) as
\[
    R_{\mathrm{com}} = \frac{d \cdot K}{\text{nonzero entries of }
    \hat{\mathbf{V}}_{K,w}^{(\ell)}},
\]
where nonzero entries are those with magnitude above $10^{-10}$. The quantity $d \cdot K$ is the number of scalar coefficients transmitted per machine under the method of Fan et al., since each local eigenvector matrix $\hat{\mathbf{V}}_K^{(\ell)} \in \mathbb{R}^{d \times K}$ is transmitted in full. That is, $R_{\mathrm{com}}$ measures how many more coefficients the method of Fan et al.\ would need to transmit compared to our method, that is, how much cheaper, in terms of communication, our method is. By construction, $R_{\mathrm{com}} \geq 1$, with larger values indicating lower communication cost relative to Fan et al.: the larger the ratio, the cheaper our method is.

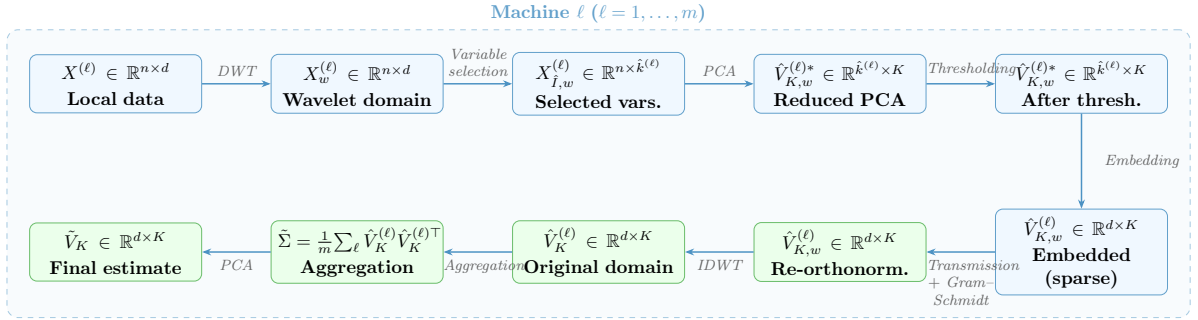
\begin{figure}[H]
\centering
\resizebox{\textwidth}{!}{%
\begin{tikzpicture}[
    node distance = 0.5cm and 1.3cm,
    caixa/.style    = {rectangle, rounded corners=4pt, draw=azul!70,
                       fill=fundocaixa, minimum width=3.2cm, minimum height=1.1cm,
                       text width=3.0cm, align=center, font=\small},
    servidor/.style = {rectangle, rounded corners=4pt, draw=verde!70,
                       fill=green!8,   minimum width=3.2cm, minimum height=1.1cm,
                       text width=3.0cm, align=center, font=\small},
    seta/.style     = {-{Stealth[length=5pt]}, thick, azul!80},
    rot/.style      = {font=\scriptsize\itshape, text=cinza,
                       align=center, text width=1.3cm},
  ]

  \node[caixa] (X)
    {$X^{(\ell)} \in \mathbb{R}^{n \times d}$\\[2pt]\textbf{Local data}};
  \node[caixa, right=1.3cm of X] (Xw)
    {$X_w^{(\ell)} \in \mathbb{R}^{n \times d}$\\[2pt]\textbf{Wavelet domain}};
  \node[caixa, right=1.3cm of Xw] (XI)
    {$X_{\hat{I},w}^{(\ell)} \in \mathbb{R}^{n \times \hat{k}^{(\ell)}}$\\[2pt]\textbf{Selected vars.}};
  \node[caixa, right=1.3cm of XI] (Vkw)
    {$\hat{V}_{K,w}^{(\ell)*} \in \mathbb{R}^{\hat{k}^{(\ell)} \times K}$\\[2pt]\textbf{Reduced PCA}};
  \node[caixa, right=1.3cm of Vkw] (Vkt)
    {$\hat{V}_{K,w}^{(\ell)*} \in \mathbb{R}^{\hat{k}^{(\ell)} \times K}$\\[2pt]\textbf{After thresh.}};

  \node[caixa, below=1.8cm of Vkt] (Vks)
    {$\hat{V}_{K,w}^{(\ell)} \in \mathbb{R}^{d \times K}$\\[2pt]\textbf{Embedded (sparse)}};
  \node[servidor, left=1.3cm of Vks] (Vreorto)
    {$\hat{V}_{K,w}^{(\ell)} \in \mathbb{R}^{d \times K}$\\[2pt]\textbf{Re-orthonorm.}};
  \node[servidor, left=1.3cm of Vreorto] (Vkfull)
    {$\hat{V}_K^{(\ell)} \in \mathbb{R}^{d \times K}$\\[2pt]\textbf{Original domain}};
  \node[servidor, left=1.3cm of Vkfull] (Sigmabar)
    {$\tilde{\Sigma} = \tfrac{1}{m}\!\sum_\ell
      \hat{V}_K^{(\ell)}\hat{V}_K^{(\ell)\top}$\\[2pt]\textbf{Aggregation}};
  \node[servidor, left=1.3cm of Sigmabar] (Vtilde)
    {$\tilde{V}_K \in \mathbb{R}^{d \times K}$\\[2pt]\textbf{Final estimate}};

  \draw[seta] (X)   -- node[above, rot] {DWT}                 (Xw);
  \draw[seta] (Xw)  -- node[above, rot] {Variable\\selection} (XI);
  \draw[seta] (XI)  -- node[above, rot] {PCA}                 (Vkw);
  \draw[seta] (Vkw) -- node[above, rot] {Thresholding}  (Vkt);
  \draw[seta] (Vkt) -- node[right, rot, text width=2.0cm]
    {Embedding} (Vks);
  \draw[seta] (Vks)      -- node[below, rot] {Transmission\\$+$ Gram--Schmidt} (Vreorto);
  \draw[seta] (Vreorto)  -- node[below, rot] {IDWT}            (Vkfull);
  \draw[seta] (Vkfull)   -- node[below, rot] {Aggregation}     (Sigmabar);
  \draw[seta] (Sigmabar) -- node[below, rot] {PCA}             (Vtilde);

  \begin{scope}[on background layer]
    \node[draw=azul!40, dashed, rounded corners=6pt, fill=azul!3,
          fit=(X)(Xw)(XI)(Vkw)(Vkt)(Vks),
          label={[font=\small\bfseries, azul!70]above:%
                 Machine $\ell$\;($\ell = 1,\ldots,m$)},
          inner sep=12pt] {};
  \end{scope}

\end{tikzpicture}%
}
\caption{Pipeline of the proposed algorithm. Blue boxes: local processing on each machine~$\ell$. Green boxes: central server. $\hat{k}^{(\ell)} = |\hat{I}^{(\ell)}|$ is the number of selected wavelet coordinates on machine $\ell$. The sparse matrices $\hat{\mathbf{V}}_{K,w}^{(\ell)}$, one per machine $\ell = 1,\ldots,m$, are the only objects transmitted; the server re-orthonormalizes via Gram-Schmidt, applies the IDWT, aggregates, and extracts the final top-$K$ eigenspace.}
\label{fig:pipeline}
\end{figure}

\section{Results and Discussion}

\subsection{Effect of the number of machines}

Before presenting the extended results across dimension that follow in Section~\ref{sec:extended-results}, we report a preliminary set of results exploring the effect of the number of machines $m$. This experiment follows the same simulation design described in Section~2.4, but with a coarser dimension grid and using only Hard thresholding, with an empirically chosen tuning parameter $c=0.6$ (rather than the universal threshold $c=1$ used elsewhere in this work). Specifically, samples are generated as in Section~2.4, with signal strengths $\lambda \in \{25, 50\}$, number of machines $m \in \{5, 10, 20\}$, dimensions $d \in \{50, 100, 200, 400, 800\}$, and $n=500$ samples per machine; all reported values are medians over $N_{MC}=100$ paired Monte Carlo replications.

\begin{table}[H]
\centering
\caption{Statistical error and communication ratio for Hard thresholding with $c=0.6$; bold indicates the lower estimation error between Fan et al.\ (2019) and the proposed method, in each row.}
\label{tab:legacy}
\small
\begin{tabular}{cc ccc ccc}
\toprule
& & \multicolumn{3}{c}{$\lambda = 50$} & \multicolumn{3}{c}{$\lambda = 25$} \\
\cmidrule(lr){3-5} \cmidrule(lr){6-8}
$m$ & $d$ & $\rho^{\mathrm{Fan}}$ & $\rho^{\mathrm{Wav}}$ & $R_{\mathrm{com}}$ & $\rho^{\mathrm{Fan}}$ & $\rho^{\mathrm{Wav}}$ & $R_{\mathrm{com}}$ \\
\midrule
\multirow{5}{*}{5}
& 50  & \textbf{0.079} & 0.290 & 7.28 & \textbf{0.119} & 0.193 & 6.15 \\
& 100 & \textbf{0.110} & 0.149 & 9.97 & 0.168 & \textbf{0.124} & 5.41 \\
& 200 & 0.157 & \textbf{0.098} & 7.56 & 0.242 & \textbf{0.157} & 4.44 \\
& 400 & 0.224 & \textbf{0.142} & 5.27 & 0.339 & \textbf{0.235} & 3.39 \\
& 800 & 0.317 & \textbf{0.213} & 3.76 & 0.481 & \textbf{0.353} & 3.15 \\
\addlinespace
\multirow{5}{*}{10}
& 50  & \textbf{0.055} & 0.282 & 7.19 & \textbf{0.083} & 0.197 & 6.18 \\
& 100 & \textbf{0.079} & 0.160 & 9.06 & 0.118 & \textbf{0.110} & 5.51 \\
& 200 & 0.112 & \textbf{0.088} & 7.35 & 0.170 & \textbf{0.119} & 3.94 \\
& 400 & 0.159 & \textbf{0.099} & 5.41 & 0.241 & \textbf{0.171} & 3.43 \\
& 800 & 0.225 & \textbf{0.153} & 3.75 & 0.345 & \textbf{0.257} & 2.97 \\
\addlinespace
\multirow{5}{*}{20}
& 50  & \textbf{0.038} & 0.286 & 7.21 & \textbf{0.058} & 0.189 & 6.16 \\
& 100 & \textbf{0.056} & 0.135 & 9.29 & \textbf{0.084} & 0.097 & 5.34 \\
& 200 & 0.079 & \textbf{0.072} & 7.17 & 0.121 & \textbf{0.091} & 4.09 \\
& 400 & 0.113 & \textbf{0.078} & 4.77 & 0.172 & \textbf{0.127} & 3.29 \\
& 800 & 0.160 & \textbf{0.109} & 3.90 & 0.246 & \textbf{0.183} & 3.03 \\
\bottomrule
\end{tabular}
\end{table}

The results in Table~\ref{tab:legacy} reveal the same dimensional boundary described in Section~\ref{sec:extended-results} below, already visible at this coarser resolution: for low dimensions ($d=50$, and $d=100$ in most rows), the algorithm of Fan et al.\ (2019) achieves lower estimation error, while for higher dimensions ($d=200,400,800$) the proposed wavelet-based method consistently outperforms it, across all three values of $m$ and both signal strengths. Regardless of $m$ and $d$, the proposed method transmits at least three times fewer nonzero coefficients than Fan et al.\ (2019), demonstrating a systematic communication advantage.

For $\lambda=50$, the estimation error of the proposed method decreases with $d$, reaching its lowest value at $d=200$ across all three values of $m$ -- precisely where it first overtakes Fan et al.\ (2019) -- and increases thereafter, while remaining below the error of Fan et al.\ (2019). The communication advantage peaks at $d=100$ and diminishes as $d$ grows, a transition that coincides with the minimum estimation error, suggesting a tradeoff between sparsification gain and eigenspace recovery quality as dimensionality increases. For $\lambda=25$, the pattern is similar, with one notable difference: the communication advantage is highest at $d=50$ and decreases with $d$. The effects of the reduced eigengap also appear in consistently lower values of $R_{\mathrm{com}}$ compared to $\lambda=50$, indicating that the proposed method must transmit more coefficients to estimate the eigenspace when the signal is weaker. Across all three values of $m$, increasing $m$ (for fixed $d$) reduces the estimation error of both methods, as expected from the larger total sample size $N=mn$, while leaving the qualitative dimensional crossover pattern unchanged.

\subsection{Extended results across dimension}
\label{sec:extended-results}

\begin{figure}[H]
\centering
\includegraphics[width=\textwidth]{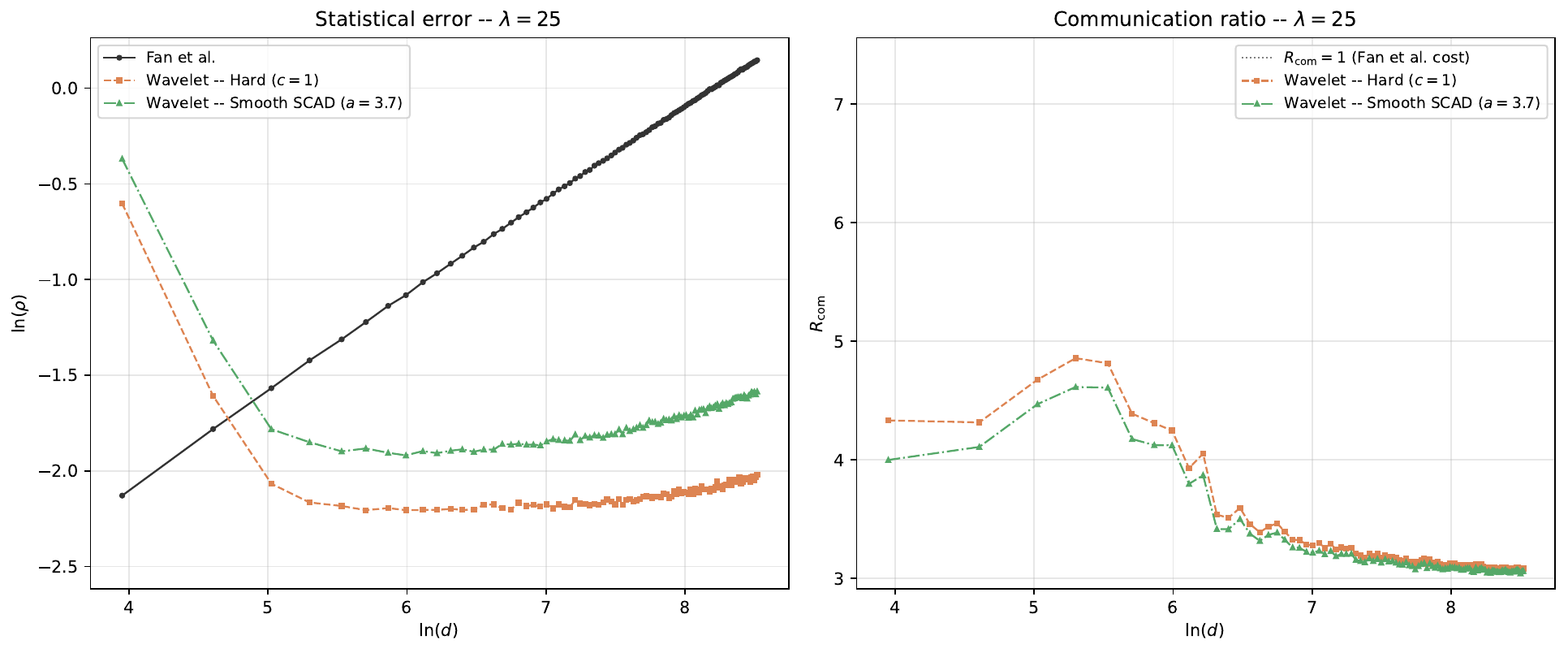}
\caption{Statistical error (left) and communication ratio (right) as a function of $\ln(d)$, for $\lambda=25$. Comparison between the algorithm of Fan et al.\ (2019) and the two thresholding variants of the proposed method, Wavelet--Hard ($c=1$) and Wavelet--Smooth SCAD ($a=3.7$).}
\label{fig:results25}
\end{figure}

\begin{figure}[H]
\centering
\includegraphics[width=\textwidth]{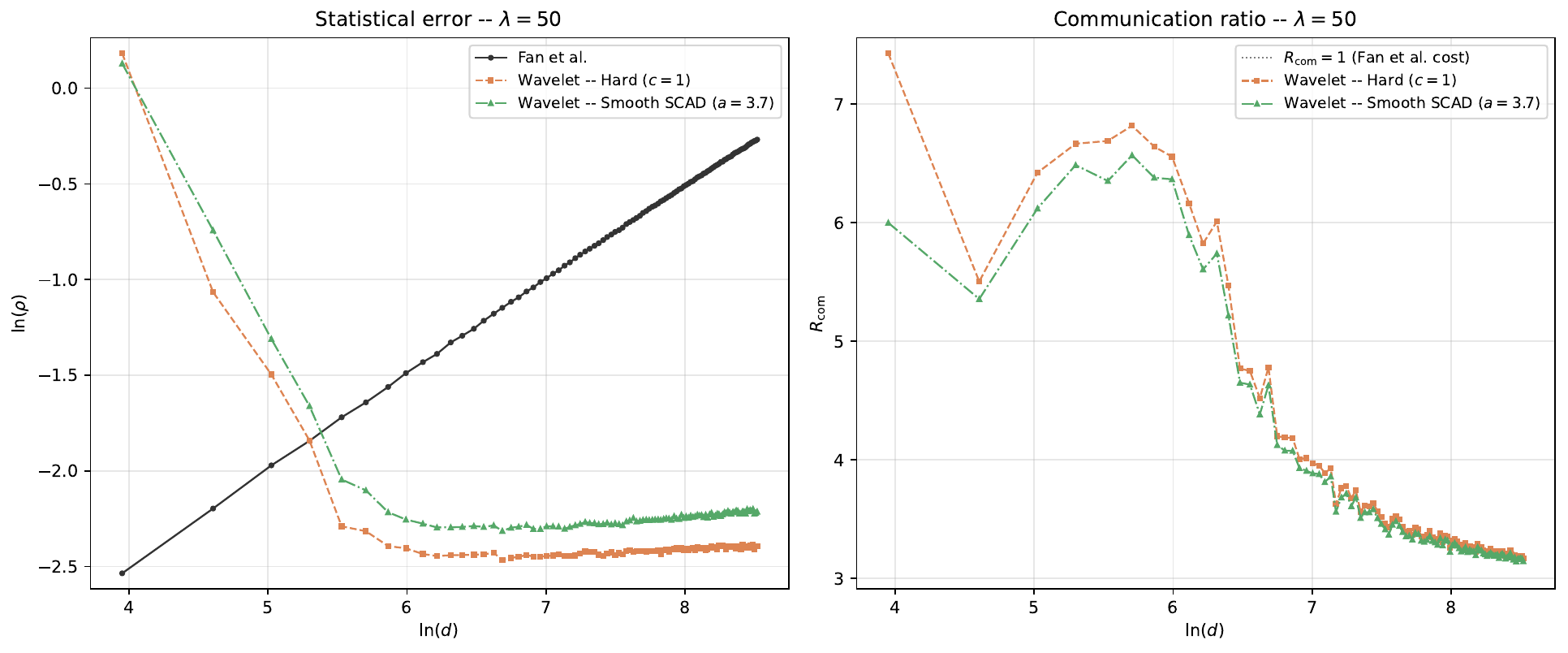}
\caption{Statistical error (left) and communication ratio (right) as a function of $\ln(d)$, for $\lambda=50$. Comparison between the algorithm of Fan et al.\ (2019) and the two thresholding variants of the proposed method, Wavelet--Hard ($c=1$) and Wavelet--Smooth SCAD ($a=3.7$).}
\label{fig:results50}
\end{figure}

Figures~\ref{fig:results25} and~\ref{fig:results50} present the results obtained for $\lambda=25$ and $\lambda=50$, respectively, following the simulation design and metrics described in the Methodology. Each figure consists of two panels: on the left, the statistical error $\ln(\rho)$ as a function of $\ln(d)$, comparing the algorithm of Fan et al.\ (2019) with the two thresholding variants of the proposed method, Wavelet--Hard ($c=1$) and Wavelet--Smooth SCAD ($a=3.7$); on the right, the communication ratio $R_{\mathrm{com}}$ as a function of $\ln(d)$, for the same two variants.

The following observations are based on the parameter configurations considered in Figures~\ref{fig:results25} and~\ref{fig:results50} and may not generalize beyond the ranges studied.

The results reveal a clear dimensional boundary separating the regimes in which each method excels. For low dimensions, the algorithm of Fan et al.\ (2019) attains lower estimation error; from $d=152$ (for $\lambda=25$) and $d=252$ (for $\lambda=50$), both thresholding variants of the proposed method -- Hard and Smooth SCAD -- consistently overtake it, maintaining this advantage up to the largest dimension considered ($d=5000$). The Hard variant's error reaches a minimum at $d=300$ ($\lambda=25$, $\rho=0.110$) and at $d=800$ ($\lambda=50$, $\rho=0.085$); the Smooth SCAD variant's minimum occurs at a comparable dimension ($d=400$ for $\lambda=25$, $\rho=0.147$; $d=800$ for $\lambda=50$, $\rho=0.099$), though at a somewhat higher error level in both cases.

As for communication cost, the ratio $R_{\mathrm{com}}$ remains always above 1 for both thresholding variants and in both scenarios, confirming that the proposed method is systematically cheaper to transmit than that of Fan et al.\ (2019). For the Hard variant, $R_{\mathrm{com}}$ starts at $4.33$ at $d=52$ ($\lambda=25$), rises to an intermediate peak at $d=200$ ($R_{\mathrm{com}}=4.86$), and decreases from there, reaching $3.08$ at $d=5000$; for $\lambda=50$, the maximum value already occurs at $d=52$ ($R_{\mathrm{com}}=7.43$), with no rising phase, decreasing thereafter to $3.17$ at $d=5000$. The Smooth SCAD variant follows a similar overall trend but transmits slightly more coefficients throughout the range: at $d=52$, $R_{\mathrm{com}}=4.00$ ($\lambda=25$) and $6.00$ ($\lambda=50$), against $4.33$ and $7.43$ for Hard; by $d=5000$ the two variants nearly coincide ($R_{\mathrm{com}} \approx 3.06$--$3.17$ for both).

Comparing the two thresholding rules directly, both perform well in this setting: Smooth SCAD attains error and communication levels close to those of Hard thresholding throughout the range studied, with Hard achieving a modest additional advantage -- lower estimation error at all 100 sampled dimensions for $\lambda=25$ and at 99 of 100 for $\lambda=50$, and a slightly higher (cheaper) communication ratio at all 100 sampled dimensions in both scenarios. It is worth noting that, while Hard thresholding uses the classical universal threshold ($c=1$), the Smooth SCAD threshold here follows a fixed heuristic rather than a fully data-driven criterion: $c(a) = \sqrt{(a-2)/(a-1)}$ with $a=3.7$, following Kulkarni et al.\ (2026, Eqs.\ 37--39). Kulkarni et al.\ (2026) also present data-driven approaches to threshold selection, which could potentially narrow this gap further, or even reverse it; we leave this investigation for future work. After reaching their respective minima, both variants' errors grow again with $d$, but the Hard variant grows more slowly than Smooth SCAD in both scenarios: $20.2\%$ versus $39.7\%$ for $\lambda=25$, and $7.3\%$ versus $10.7\%$ for $\lambda=50$ (relative growth between each variant's own minimum and $d=5000$).

In both scenarios, after reaching the minimum estimation error, the Hard variant's error grows again with $d$, but much more slowly than the error of Fan et al.\ (2019): between the Hard variant's minimum-error point and $d=5000$, the error of Fan et al.\ (2019) grows by $293.0\%$ ($\lambda=25$) and $140.9\%$ ($\lambda=50$), against only $20.2\%$ and $7.3\%$, respectively. This slower growth, however, comes at a cost: the number of transmitted coefficients also grows with $d$, which is reflected in the reduction of $R_{\mathrm{com}}$, our communication advantage over Fan et al.\ (2019), as $d$ increases.

The effect of the difference between the signal intensities considered is also evident in this post-minimum growth for the Hard variant: as seen above, it is much slower for $\lambda=50$ ($7.3\%$) than for $\lambda=25$ ($20.2\%$), a difference that is quite visible in the corresponding plots. Simultaneously, the $R_{\mathrm{com}}$ curve for $\lambda=50$ is shifted upward relative to the curve for $\lambda=25$ throughout practically the entire range of $d$ studied, for both thresholding variants. That is, with a stronger signal, the proposed method simultaneously attains a lower estimation error and a smaller number of transmitted coefficients, regardless of the thresholding rule employed.

\section{Conclusion}

The incorporation of wavelet-based sparsification into the distributed PCA of Fan et al.\ (2019) proved advantageous precisely in the regime where classical distributed PCA is least efficient: high dimensions comparable to the local sample size. In this regime ($d \geq 152$ for $\lambda=25$ and $d \geq 252$ for $\lambda=50$, in the simulation considered), the proposed method -- under either thresholding rule -- not only systematically reduces communication cost, with $R_{\mathrm{com}}$ always above 1, but also produces lower-error estimates, reversing the result observed at low dimensions. This behavior is consistent with the theory of Johnstone \& Lu (2009): by restricting PCA to a sparse subset of relevant coordinates in the wavelet domain, one avoids the accumulation of error across uninformative dimensions that compromises classical PCA in high dimension. Between the two thresholding rules considered, both Hard thresholding with the universal threshold ($c=1$) and Smooth SCAD ($a=3.7$) performed well in this setting, with Hard achieving a modest additional advantage in both estimation error and communication cost. Since the Smooth SCAD threshold was set via a fixed heuristic rather than a fully data-driven criterion, exploring the data-driven approaches presented by Kulkarni et al.\ (2026) is a promising direction for future refinement of the algorithm. More broadly, the results also expose important \textit{tradeoffs} between sparsification and eigenspace recovery quality, and between signal intensity and communication advantage that constitute natural directions for the theoretical and empirical refinement of the algorithm.

\section*{Acknowledgments}

This study was financed by the S\~ao Paulo Research Foundation (FAPESP), Brazil. Process Number \#2023/02538-0 and Number \#2025/21250-2.

\section*{References}

\begin{footnotesize}
\noindent ANSEL, J.; YANG, E.; HE, H.; GIMELSHEIN, N.; JAIN, A.; VOZNESENSKY, M.; et al. \textbf{PyTorch 2: Faster Machine Learning Through Dynamic Python Bytecode Transformation and Graph Compilation.} Proceedings of the 29th ACM International Conference on Architectural Support for Programming Languages and Operating Systems (ASPLOS '24), Volume 2. ACM, 2024. DOI: 10.1145/3620665.3640366.

\vspace{0.2cm}
\noindent CHEN, X.; LEE, J. D.; LI, H.; YANG, Y. \textbf{Distributed Estimation for Principal Component Analysis: An Enlarged Eigenspace Analysis.} Journal of the American Statistical Association, 117(540):1775--1786, 2022. DOI: 10.1080/01621459.2021.1886937.

\vspace{0.2cm}
\noindent FAN, J.; WANG, D.; WANG, K.; ZHU, Z. \textbf{Distributed Estimation of Principal Eigenspaces.} Annals of Statistics, 47(6):3009--3031, 2019. DOI: 10.1214/18-AOS1713.

\vspace{0.2cm}
\noindent FAN, J.; LI, R.; ZHANG, C.-H.; ZOU, H. \textbf{Statistical Foundations of Data Science.} 1st ed. Chapman and Hall/CRC, 2020. DOI: 10.1201/9780429096280.

\vspace{0.2cm}
\noindent FAN, J.; LI, R. \textbf{Variable Selection via Nonconcave Penalized Likelihood and Its Oracle Properties.} Journal of the American Statistical Association, 96(456):1348--1360, 2001. DOI: 10.1198/016214501753382273.

\vspace{0.2cm}
\noindent HARRIS, C. R.; MILLMAN, K. J.; VAN DER WALT, S. J.; GOMMERS, R.; VIRTANEN, P.; COURNAPEAU, D.; et al. \textbf{Array Programming with NumPy.} Nature, 585(7825):357--362, 2020. DOI: 10.1038/s41586-020-2649-2.

\vspace{0.2cm}
\noindent HUNTER, J. D. \textbf{Matplotlib: A 2D Graphics Environment.} Computing in Science \& Engineering, 9(3):90--95, 2007. DOI: 10.1109/MCSE.2007.55.

\vspace{0.2cm}
\noindent JOHNSON, R. A.; WICHERN, D. W. \textbf{Applied Multivariate Statistical Analysis.} 6th ed. Upper Saddle River: Pearson Prentice Hall, 2007.

\vspace{0.2cm}
\noindent JOHNSTONE, I. M. \textbf{On the Distribution of the Largest Eigenvalue in Principal Components Analysis.} Annals of Statistics, 29(2):295--327, 2001. DOI: 10.1214/aos/1009210544.

\vspace{0.2cm}
\noindent JOHNSTONE, I. M.; LU, A. Y. \textbf{On Consistency and Sparsity for Principal Components Analysis in High Dimensions.} Journal of the American Statistical Association, 104(486):682--693, 2009. DOI: 10.1198/jasa.2009.0121.

\vspace{0.2cm}
\noindent KULKARNI, R.; PINHEIRO, A.; VIDAKOVIC, B.; ATTO, A. M. \textbf{Smooth SCAD: A Raised Cosine Thresholding Rule for Wavelet Denoising.} Mathematics, 14(5):787, 2026. DOI: 10.3390/math14050787.

\vspace{0.2cm}
\noindent LEE, G. R.; GOMMERS, R.; WASILEWSKI, F.; WOHLFAHRT, K.; O'LEARY, A. \textbf{PyWavelets: A Python Package for Wavelet Analysis.} Journal of Open Source Software, 4(36):1237, 2019. DOI: 10.21105/joss.01237.

\vspace{0.2cm}
\noindent MCKINNEY, W. \textbf{Data Structures for Statistical Computing in Python.} Proceedings of the 9th Python in Science Conference, pp. 56--61, 2010. DOI: 10.25080/Majora-92bf1922-00a.

\vspace{0.2cm}
\noindent MORETTIN, P. A.; PINHEIRO, A.; VIDAKOVIC, B. \textbf{Wavelets in Functional Data Analysis.} Cham: Springer, 2017. DOI: 10.1007/978-3-319-59623-5.

\vspace{0.2cm}
\noindent OPPENHEIM, A. V.; WILLSKY, A. S.; NAWAB, S. H. \textbf{Signals and Systems.} 2nd ed. Upper Saddle River: Prentice Hall, 1997.

\vspace{0.2cm}
{\sloppy\noindent VIDAKOVIC, B. \textbf{Statistical Modeling by Wavelets.} New York: Wiley-Interscience, 1999. DOI: 10.1002/9780470317020.\par}

\vspace{0.2cm}
\noindent VIRTANEN, P.; GOMMERS, R.; OLIPHANT, T. E.; HABERLAND, M.; REDDY, T.; COURNAPEAU, D.; et al. \textbf{SciPy 1.0: Fundamental Algorithms for Scientific Computing in Python.} Nature Methods, 17(3):261--272, 2020. DOI: 10.1038/s41592-019-0686-2.
\end{footnotesize}

\end{document}